# Sub-critical model of damage evolution as a phase transition

S.G. Abaimov

E-mail: sgabaimov@gmail.com

**PACS.** 62.20.M- Structural failure of materials - 89.75.-k Complex systems - 05. Statistical physics, thermodynamics, and nonlinear dynamical systems

**Abstract.** Many studies investigated the application of statistical mechanics to damage phenomena. However, so far the association of damage with statistical mechanics is far from completely developed. One of the most successful approaches maps the strength elements onto the spins of a lattice. This allows applying the formalism extensively developed for spin systems to damage phenomena. To improve the model the interactions of spins at the crack tips are introduced. Further, statistical mechanics of the model is developed for different external boundary constraints.

## 1. Introduction

An analogy between damage mechanics and phase transitions has been discussed by many authors. One of the earliest approaches was suggested by Smalley et al. [1]. In their study the authors applied the formalism of the theory of renormalization groups to damage phenomena. Unfortunately, this promising approach has later found a little continuation in the literature. Rundle and Klein [2] introduced a free energy potential for systems with damage and investigated the process of crack growth as a nucleation process in the vicinity of a spinodal. Blumberg Selinger et al. [3] investigated rupture processes in a fiber-bundle model and in a model of a two-dimensional network and associated a fracture with an approach to a spinodal. Spinodal power-law behavior of damage as a first-order transition was investigated by Zapperi et al. [4]. Also this study investigated dynamical behavior of avalanches and found that avalanches in the vicinity of a spinodal exhibit power-law scaling dependences.

The first approach to differentiate thermal and topological characteristics of damage occurrence was suggested by Pride and Toussaint [5] and Toussaint and Pride [6]. In these studies the concept of topological entropy was introduced which later leaded the author of this manuscript to develop a rigorous formalism of statistical mechanics for the topological occurrence of damage phenomena [7, 8].

However, so far the association of damage with statistical mechanics is far from completely developed. The complexity of the description of damage based on the theory of elasticity makes it difficult to apply statistical mechanics directly to the processes of crack evolution. The situation is analogous to the case of spontaneous magnetization in magnetic systems where the complexity of quantum exchange effects among atoms and particles in crystal structures disguises the nature of the problem. The first step to overcome this difficulty is the derivation of a simplified model to reveal the basic characteristics of the behavior. This has been done in the theory of magnetic systems with the introduction of the Ising model. In spite of its simplicity in comparison with real systems, the Ising model describes the principal characteristics of behavior and has become a foundation for the theory of phase transitions for magnetic systems. Therefore we need a similar simplified, illustrative model that reflects the basic principles of damage occurrence in solids.

The first important attempt to describe the phenomenon of damage with a spin model was made by Blumberg Selinger *et al.* [3]. They introduced a Hamiltonian model that mapped the strength elements of a fiber-bundle model onto the spins on a lattice. Damage phenomena were modeled by the equilibrium detailed balance of broken and intact fibers. Broken fibers were allowed to become intact again which corresponds to systems with perfect healing (e.g., gels) or to the sub-critical micro-initiation of damage when the level of damage is much less than the critical threshold and small cracks can be considered as reversible. No interactions of fibers were assumed. Later the same model was modified by Virgilii *et al.* [9]. They introduced interactions among spins. However, spins in their model interact only by means of the global (mean-field) stress redistribution.

To enrich the model behavior we introduce a modified Hamiltonian that includes spin-spin interactions between nearest spin neighbors. This modification introduces another phase into the system.

In Section 2 we describe the model. In Section 3 we introduce the nearest-neighbor fiber interactions at crack tips and consider the case of the constant strain. For the same interactions the case of the constant stress is considered in Section 4.

## 2. Model

Following Blumberg Selinger *et al.* [3] and Virgilii *et al.* [9], we first consider a fiber-bundle model with the global, or mean-field, load sharing. Two absolutely rigid surfaces are connected by a bundle of fibers. Each undisturbed fiber has length $l_0$, cross-sectional area $s_0$, and volume $v_0 = l_0 \, s_0$. The total number of fibers in the model is $N$, and the fibers can be intact or broken. We denote the fraction of intact fibers as $L$. Then the damage variable $D = 1 - L$ represents the fraction of broken fibers. Both variables represent order parameters distinguishing the opposite phases. As a microstate of the system we consider a particular realization {S} of intact and broken fibers on the lattice. As a macrostate [L] we consider the union of all microstates corresponding to the given value of $L$.

The behavior of individual intact fibers is assumed to be perfectly elastic with the Young modulus, $E$, being the same for all fibers. We assume that the external force $F_{ext}$ is applied to the total surface of the model $Ns_0$. This permits the introduction of a 'virtual' external stress as the ratio of the external force to the total surface of the model $\sigma_{ext} = F_{ext} / Ns_0$. An external observer, who would see the model as a 'black box' and would not know about the damage inside, would assume that this is the stress acting in the system. As some fibers are broken, the real stress on the intact fibers is higher than $\sigma_{ext}$ and increases with the damage propagation as $\sigma_f = F_{ext}/(LNs_0) = \sigma_{ext}/L$. We assume that all intact fibers have the same strain $\varepsilon = \sigma_f / E$.

Let index *i* enumerate the fibers. Following Blumberg Selinger *et al.* [3], we introduce the 'spin' variables $S_i$ with unity value of spin for intact fibers $S_i = 1$ and zero value for broken fibers $S_i = 0$. Then for a microstate {S} the order parameter $L_{\{S\}}$ of this microstate is the averaged value of spins $L_{\{S\}} = N^{-1} \sum_{i=1}^{N} S_i$.

We further assume that there is thermodynamic equilibrium among intact and broken fibers, and that the fibers can switch their states between these two options due to thermal fluctuations. To become broken a fiber has to consume energy $\gamma$ to form two rupture surfaces. At the same time the fiber emits its elastic energy $\frac{1}{2} E v_0 \varepsilon^2$. To become intact we assume that the fiber must emit its surface energy $\gamma$ and to consume elastic energy $\frac{1}{2} E v_0 \varepsilon^2$, where $\varepsilon$ is the strain of intact fibers at the same time [3]. Here we assume that there is healing in the model and to become healed a fiber has to reproduce the strain of the surrounding intact fibers.

## 2.1 External boundary constraints of constant temperature *T* and constant strain $\varepsilon$ = const. System without interactions

A Hamiltonian for this system can be introduced [3] as:

$$H_{\{S\}} = \sum_{i:S_i=1} \frac{1}{2} E v_0 \varepsilon^2 + \sum_{i:S_i=0} \gamma = N\gamma - \sum_{i=1}^{N} \left( \gamma - \frac{1}{2} E v_0 \varepsilon^2 \right) S_i, \qquad (1)$$

where the first sum is over all intact fibers, the second sum is over all broken fibers, and the third sum is over all *N* fibers. $N\gamma$ is the constant shift of energy and will not be included in further formulae. Also we see that the Hamiltonian depends only on the order parameter $L$ and is the same for all microstates $\{S\}$ corresponding to the given $L$:

$$H_{\{S\}}(L) = -NL\left(\gamma - \frac{1}{2}E\upsilon_0\varepsilon^2\right) \qquad (2)$$

as expected for the absence of interactions. Strain $\varepsilon$ here is constant and therefore the system is a typical two-level system where spins interact only with the external field $B_{ext} = \gamma - \frac{1}{2}E\upsilon_0\varepsilon^2$ and do not interact with each other. The partition function is

$$Z = \sum_{\{S\}} e^{-\beta H_{\{S\}}} = \left(\sum_{S=0,1} e^{\beta\left(\gamma - \frac{1}{2}E\upsilon_0\varepsilon^2\right)S}\right)^N = \left(1 + e^{\beta\left(\gamma - \frac{1}{2}E\upsilon_0\varepsilon^2\right)}\right)^N, \qquad (3)$$

and the equilibrium solution is

$$<L> = \frac{e^{\beta\left(\gamma - \frac{1}{2}E\upsilon_0\varepsilon^2\right)}}{1 + e^{\beta\left(\gamma - \frac{1}{2}E\upsilon_0\varepsilon^2\right)}}. \qquad (4)$$

By definition, the two-level system has only one, unique solution (4) and a phase transition is not possible in this system. This can be verified through derivation of the free energy potential. We will carefully look at this question for this trivial case because this technique will be employed for other, more complex models.

The Hamiltonian of the system under consideration is similar to the magnetic Ising model and can be mapped exactly on the Ising model without interactions. The free energy potential for the Ising model is the Helmholtz energy. Therefore, we hypothesize that the free energy potential for our system is also the Helmholtz energy. Our system is under the constant temperature and constant strain (constant volume) conditions. The free energy potential for these boundary conditions in statistical mechanics of elastic media is also the Helmholtz energy. Therefore, we again expect that to study behavior of the system we should investigate the Helmholtz free energy for non-equilibrium macrostates $[L]$.

First we find the number of microstates corresponding to the given macrostate $[L]$. This number is given by the simple combinatorial choice of $N_1 = NL$ intact fibers and $N_0 = N(1-L)$ broken fibers among all $N$ fibers:

$$g_{[L]} = \frac{N!}{N_1! N_0!} \approx_{\ln} \frac{1}{L^{NL}} \frac{1}{(1-L)^{N(1-L)}}, \tag{5}$$

where we use Stirling's formula and the symbol $\approx_{\ln}$ means that all power-law multipliers have been neglected in comparison with the exponential dependence on $N$ (which is infinite in the thermodynamic limit). Then the entropy of this non-equilibrium macrostate $[L]$ (the entropy of a system isolated on this non-equilibrium macrostate $[L]$) is

$$S_{[L]} = k_B \ln g_{[L]}. \tag{6}$$

For the Helmholtz free energy of this non-equilibrium macrostate $[L]$ we obtain

$$A_{[L]} \equiv H_{[L]} - TS_{[L]} = -k_B T \ln\left(g_{[L]} e^{-\beta H_{[L]}}\right) = -k_B T \ln Z_{[L]}, \tag{7}$$

where $Z_{[L]}$ is the partial partition function [8] only over microstates corresponding to the given macrostate $[L]$. The second derivative of the Helmholtz free energy is

$$\frac{\partial^2 A_{[L]}}{\partial L^2} = \frac{Nk_B T}{L(1-L)}. \tag{8}$$

This quantity is positive for all possible values of $L$: $0 \leq L \leq 1$ (for all macrostates $[L]$) and therefore the free energy potential has only one minimum.

## 2.2 External boundary constraints of constant temperature $T$ and constant stress $\sigma_{ext}$ = const. System with mean-field interactions

The case of constant external stress has been considered by Virgilii *et al.* [9]. We can use the previous Eq. (1) for the Hamiltonian, but substitute variable strain $\varepsilon$ as $\varepsilon = \sigma_{ext}/(EL)$:

$$H_{\{S\}} = -\sum_{i=1}^{N} \left( \gamma - \frac{1}{2} E \upsilon_0 \left\{ \frac{\sigma_{ext}}{E \frac{1}{N} \sum_{j=1}^{N} S_j} \right\}^2 \right) S_i. \tag{9}$$

Again we see that the Hamiltonian depends only on the order parameter $L$:

$$H_{[L]}(L) = -NL \left( \gamma - \frac{1}{2} E \upsilon_0 \left\{ \frac{\sigma_{ext}}{EL} \right\}^2 \right). \tag{10}$$

For the partition function we have

$$\zeta = \sum_{L=0:\Delta L=1/N}^{1} g_{[L]} e^{-\beta(H_{[L]} - N\upsilon_0 \sigma \varepsilon_{[L]})}, \tag{11}$$

where the sum is assumed to be over the discrete values of $L$ with the step $\Delta N_1 = 1$ corresponding to one fiber flip. In Eq. (11) we can exchange the sum $\sum_{L=0: \Delta L=1/N}^{1}$ by the integral $\int_0^1 \frac{dL}{1/N}$. The expression for the partition function then becomes:

$$\zeta = \int_0^1 \frac{dL}{1/N} e^{Nf(L)}, \tag{12}$$

where $f(L)$ is given by

$$f(L) = -L \ln L - (1-L) \ln(1-L) + \beta\gamma L + \beta \frac{1}{2} E \upsilon_0 \left(\frac{\sigma_{ext}}{E}\right)^2 \frac{1}{L}. \tag{13}$$

To evaluate this integral we need to use the saddle point method. Again neglecting all power-law multipliers in comparison with the exponential dependence on $N$ we obtain

$$\zeta \approx_{\ln} e^{Nf(L_0)} \text{ where } L_0 \text{ is a solution of } \frac{\partial f}{\partial L}(L_0) = 0. \tag{14}$$

It is easy to prove that $L_0$ is, in fact, the equilibrium value of the order parameter $<L>$. Therefore, to determine a presence of a possible phase transition in the system we need to find whether the transcendental equation $\partial f / \partial L(L_0) = 0$ has one or more solutions. This is straightforward and one can find that the solution is not unique and there is a phase transition corresponding to damage instability. But better is again to look at the behavior of the free energy potential.

For the damage model we have now constant temperature and constant stress (constant pressure) as boundary constraints. In statistical mechanics of elastic systems the free energy potential for these conditions is the Gibbs energy. For a macrostate [L] we have $G_{[L]} = -k_B T N f(L)$. After simple algebra we obtain

$$\frac{\partial^2 G_{[L]}}{\partial L^2} = \frac{Nk_B T}{L(1-L)} - NE\upsilon_0 \left(\frac{\sigma_{ext}}{E}\right)^2 \frac{1}{L^3}. \tag{21}$$

Because its second derivative can be positive as well as negative the Gibbs free energy can have two minima. Therefore, there is a phase transition in the system that would correspond to instability of damage.

## 3. External boundary constraints of constant temperature *T* and constant strain *ε* = const. System with nearest-neighbor interactions

The systems considered in the previous two sections do not include interactions of fibers or include them in a mean-field approximation. In this paper we introduce the direct spin-spin interactions of nearest neighbors which allow us to enrich the behavior of the model.

The most unstable parts of a solid with damage are the tips of the cracks. Often in fracture mechanics the elastic solution for the stress distribution has a singularity at this tip. The reason is that the atomic bonds at the crack tip are frustrated between two boundary constraints: an open crack at one side and the absence of an opening at the other side of the crack. Therefore for the spin model it is also reasonable to assume that spins at the crack tips are frustrated in a similar fashion. In other words, intact and broken fibers that are nearest neighbors must have higher energy than if both fibers are intact or both fibers are broken. This will make the crack tips behave as the most unstable points of damage growth in the model.

So, we assume that neighboring intact $S_i = 1$ and broken $S_j = 0$ fibers have excess of energy $J$. Neighboring intact fibers do not have additional energy, nor do neighboring broken fibers. Note that the construction $\{S_i(1-S_j) + S_j(1-S_i)\}$ equals to unity only when $S_i$ and $S_j$ are different and equals to zero otherwise. Therefore we can write the Hamiltonian for the system as

$$H_{\{S\}} = -\sum_{i=1}^{N}\left(\gamma - \frac{1}{2}E\upsilon_0\varepsilon^2\right)S_i + J\sum_{<i,j>_{n.n.}}\{S_i(1-S_j) + S_j(1-S_i)\}, \quad (22)$$

where the sum $\sum_{<i,j>_{n.n.}}$ means the sum over nearest neighbor pairs. We can rewrite this expression as

$$H_{\{S\}} = -\left(\gamma - \frac{1}{2}E\upsilon_0\varepsilon^2\right)\sum_{i=1}^{N}S_i + J\sum_{<i,j>_{n.n.}}\{S_i - 2S_iS_j + S_j\}. \quad (23)$$

As $\sum_{<i,j>_{n.n.}}S_i = \frac{1}{2}\sum_{i=1}^{N}qS_i$ where $q$ is the coordination number of the lattice we obtain

$$H_{\{S\}} = -\left(\gamma - \frac{1}{2}E\upsilon_0\varepsilon^2 - Jq\right)\sum_{i=1}^{N}S_i - 2J\sum_{<i,j>_{n.n.}}S_iS_j. \quad (24)$$

To obtain an analytical solution we have to use a mean-field approximation $\frac{1}{Nq/2}\sum_{<i,j>_{n.n.}} S_i S_j = \left(\frac{1}{N}\sum_{i=1}^{N} S_i\right)^2 = L^2$. Then the Hamiltonian again depends only on the order parameter $L$

$$H_{[L]}(L) = -NL\left(\gamma - \frac{1}{2}E\upsilon_0\varepsilon^2 - Jq\right) - NqJL^2. \tag{25}$$

For the partition function we obtain

$$Z = e^{Nf(L_0)} \text{ where } L_0 = <L> \text{ is a solution of } \frac{\partial f}{\partial L}(L_0) = 0 \tag{26}$$

and the function $f(L)$ is given by

$$f(L) = -L\ln L - (1-L)\ln(1-L) + \beta L\left(\gamma - \frac{1}{2}E\upsilon_0\varepsilon^2 - Jq\right) + \beta qJL^2. \tag{27}$$

Investigation shows that the transcendental equation obtained for $L_0$ under some particular values of the external constraints has two independent solutions. But again, it is much simpler to investigate the behavior of the Helmholtz energy as a free energy potential. For non-equilibrium macrostates $[L]$ we have $A_{[L]} = -k_B TNf(L)$ and

$$\frac{\partial^2 A_{[L]}}{\partial L^2} = \frac{Nk_B T}{L(1-L)} - 2NqJ. \tag{28}$$

The second derivative of the free energy potential can be both negative and positive. This means that for high values of $q$ and $J$ the second minimum of the potential appears. The stronger interactions overwhelm the destructing effect of temperature and create two opposite phases with high and low values of damage. This is visualized simply by mapping the system under consideration on the Ising model. Indeed, if we introduce new spin variables $\sigma_i = 1 - 2S_i$ (as a traditional mapping of the lattice gas model $S_i = 0,1$ on the Ising model $\sigma_i = \pm 1$), then for the Hamiltonian (24) we obtain

$$H_{\{\sigma\}} = -\frac{N}{2}\left(\gamma - \frac{1}{2}E\upsilon_0\varepsilon^2 - Jq\right) + \frac{1}{2}\left(\gamma - \frac{1}{2}E\upsilon_0\varepsilon^2 - Jq\right)\sum_{i=1}^{N}\sigma_i - \frac{J}{2}\sum_{<i,j>_{n.n.}}\left(1 - \sigma_i - \sigma_j + \sigma_i\sigma_j\right). \quad (29)$$

Recalling again that $\sum_{<i,j>_{n.n.}}\sigma_i = \frac{1}{2}\sum_{i=1}^{N}q\sigma_i$ and $\sum_{<i,j>_{n.n.}}1 = Nq/2$, we obtain

$$H_{\{\sigma\}} = -\frac{N}{2}\left(\gamma - \frac{1}{2}E\upsilon_0\varepsilon^2 - \frac{Jq}{2}\right) + \frac{1}{2}\left(\gamma - \frac{1}{2}E\upsilon_0\varepsilon^2\right)\sum_{i=1}^{N}\sigma_i - \frac{J}{2}\sum_{<i,j>_{n.n.}}\sigma_i\sigma_j. \quad (30)$$

This is the Hamiltonian of the classical Ising model with the magnetic field $B_{ext} = -\frac{1}{2}\left(\gamma - \frac{1}{2}E\upsilon_0\varepsilon^2\right)$ and interaction constant $J/2$. Therefore, for lower temperatures below critical, the damage model exhibits the presence of two phases and the phase transition between them corresponds to the brittle fracture. For higher temperatures above critical there is one unique phase and damage is accumulated in the regime without spontaneous instabilities.

## 4. External boundary constraints of constant temperature $T$ and constant stress $\sigma_{ext}$ = const. System with nearest-neighbor interactions

We can introduce a similar system with interactions for the external constraint of constant stress. The Hamiltonian for the system is

$$H_{\{S\}} = -\sum_{i=1}^{N}\left(\gamma - \frac{1}{2}E\upsilon_0\left(\frac{\sigma_{ext}}{E\frac{1}{N}\sum_{j=1}^{N}S_j}\right)^2\right)S_i + J\sum_{<i,j>_{n.n.}}\{S_i(1-S_j) + S_j(1-S_i)\} \text{ or} \quad (31)$$

$$H_{\{S\}} = -\sum_{i=1}^{N}\left(\gamma - \frac{1}{2}E\upsilon_0\left(\frac{\sigma_{ext}}{E\frac{1}{N}\sum_{j=1}^{N}S_j}\right)^2 - Jq\right)S_i - 2J\sum_{<i,j>_{n.n.}}S_iS_j. \quad (32)$$

To obtain an analytical solution we have to use the meanfield approximation

$$\frac{1}{Nq/2}\sum_{<i,j>_{n.n.}}S_iS_j = \left(\frac{1}{N}\sum_{i=1}^{N}S_i\right)^2 = L^2. \text{ Then for the Hamiltonian we obtain}$$

$$H_{[L]} = -NL\left(\gamma - \frac{1}{2}E\upsilon_0\left(\frac{\sigma_{ext}}{EL}\right)^2 - Jq\right) - NqJL^2. \quad (33)$$

The partition function is

$$\zeta = e^{Nf(L_0)} \text{ where } L_0 = <L> \text{ is a solution of } \frac{\partial f}{\partial L}(L_0) = 0 \quad (34)$$

and the function $f(L)$ is given by

$$f(L) = -L\ln L - (1-L)\ln(1-L) + \beta L\left(\gamma + \frac{1}{2}E\upsilon_0\left(\frac{\sigma_{ext}}{EL}\right)^2 - Jq\right) + \beta qJL^2. \quad (35)$$

For non-equilibrium macrostates [L] we have $G_{[L]} = -k_BTNf(L)$ and

$$\frac{\partial^2 G_{[L]}}{\partial L^2} = \frac{Nk_BT}{L(1-L)} - NE\upsilon_0\left(\frac{\sigma_{ext}}{E}\right)^2\frac{1}{L^3} - 2NqJ. \quad (36)$$

The second derivative of the free energy potential can be both negative and positive. This again results in the appearance of another minimum for high values of $q$, $J$, and $\sigma_{ext}$. The stronger interactions and stress overwhelm the destructing effect of temperature and create phases with high and low values of damage. We associate this with the presence of brittle fracture. However, high temperature breaks the coexistence of phases and causes the system to follow the simple one phase solution without brittle instabilities.

## 7. Conclusions

To enrich the model behavior we include the interactions of nearest neighbors at crack tips. This introduces the instability at the crack tips as the main points of fracture propagation and results in the appearance of a phase transition in the system. We associate this phase transition with the brittle regime of fracture below the critical temperature.

## References

1. R. F. Smalley, D. L. Turcotte, and S. A. Solla, J. Geophys. Res. **90**, 1894 (1985)

2. J. B. Rundle and W. Klein, Phys. Rev. Lett. **63**, 171 (1989)

3. R. L. Blumberg Selinger, Z.-G. Wang, and W. M. Gelbart, Phys. Rev. A **43**, 4396 (1991)

4. S. Zapperi, P. Ray, H. E. Stanley, and A. Vespignani, Phys. Rev. Lett. **78**, 1408 (1997)

5.  S. R. Pride and R. Toussaint, Physica A **312**, 159 (2002)

6.  R. Toussaint and S. R. Pride, Phys. Rev. E **66**, 036135 (2002); R. Toussaint and S. R. Pride, Phys. Rev. E **66**, 036136 (2002); R. Toussaint and S. R. Pride, Phys. Rev. E **66**, 036137 (2002); R. Toussaint and S. R. Pride, Phys. Rev. E **71**, 046127 (2005)

7.  S. G. Abaimov, J. Stat. Mech., P09005 (2008)

8.  S. G. Abaimov, J. Stat. Mech., P03039 (2009)

9.  A. Virgilii, A. Petri, and S. R. Salinas, J. Stat. Mech., P04009 (2007)